\documentclass[journal]{IEEEtran}
\usepackage{amsmath,amssymb,bbm,bm,graphicx,braket,upgreek}
\usepackage[absolute]{textpos}
\usepackage{xcolor}
\usepackage[noadjust]{cite}
\usepackage[hidelinks]{hyperref}
\usepackage[capitalize]{cleveref}
\crefname{section}{Sec.}{}

\def\cF{\mathcal{F}}
\def\cP{\mathcal{P}}
\DeclareMathOperator{\Tr}{Tr}

\begin{document}
\bstctlcite{IEEEexample:BSTcontrol}
\title{Design Methodologies for Integrated Quantum Frequency Processors}

\author{Benjamin E.~Nussbaum,
Andrew J.~Pizzimenti,
Navin B.~Lingaraju,\\
Hsuan-Hao Lu, and
Joseph M.~Lukens
\thanks{B.~E.~Nussbaum is with the Department of Physics and Illinois Quantum Information Science \& Technology Center (IQUIST), University of Illinois Urbana-Champaign, Urbana, Illinois 61801, USA (e-mail: bn9@illinois.edu).}
\thanks{A.~J.~Pizzimenti is with the James C.~Wyant College of Optical Sciences, University of Arizona, Tucson, Arizona 85721, USA (e-mail: ajpizzimenti@email.arizona.edu).}
\thanks{N.~B.~Lingaraju is with SRI International, Arlington, Virginia 22209, USA (e-mail: navin.lingaraju@sri.com).}
\thanks{H.-H.~Lu and J.~M.~Lukens are with the Quantum Information Science Section, Oak Ridge National Laboratory, Oak Ridge, Tennessee 37831, USA (e-mail: luh2@ornl.gov, lukensjm@ornl.gov).}
}

{}
\maketitle

\begin{abstract}
Frequency-encoded quantum information offers intriguing opportunities for quantum communications and networking, with the quantum frequency processor paradigm---based on electro-optic phase modulators and Fourier-transform pulse shapers---providing a path for scalable construction of quantum gates.
Yet all experimental demonstrations to date have relied on discrete fiber-optic components that occupy significant physical space and impart appreciable loss.
In this article, we introduce a model for the design of quantum frequency processors comprising microring resonator-based pulse shapers and integrated phase modulators.
We estimate the performance of single and parallel frequency-bin Hadamard gates, finding high fidelity values that extend to frequency bins with relatively wide bandwidths.
By incorporating multi-order filter designs as well, we explore the limits of tight frequency spacings, a regime extremely difficult to obtain in bulk optics.
Overall, our model is general, simple to use, and extendable to other material platforms, providing a much-needed design tool for future frequency processors in integrated photonics.
\end{abstract}

\begin{IEEEkeywords}
Quantum computing, silicon photonics, optical pulse shaping, photonic integrated circuits, optical resonators, phase modulation.
\end{IEEEkeywords}

\IEEEpeerreviewmaketitle

\begin{textblock}{13.4}(1.3,15.3)\noindent\fontsize{6}{6}\selectfont\textcolor{black!30}{This manuscript has been co-authored by UT-Battelle, LLC, under contract DE-AC05-00OR22725 with the US Department of Energy (DOE). The US government retains and the publisher, by accepting the article for publication, acknowledges that the US government retains a nonexclusive, paid-up, irrevocable, worldwide license to publish or reproduce the published form of this manuscript, or allow others to do so, for US government purposes. DOE will provide public access to these results of federally sponsored research in accordance with the DOE Public Access Plan (http://energy.gov/downloads/doe-public-access-plan).} \end{textblock}

\section{Introduction}
\label{sec:intro}
Frequency encoding is emerging as a promising paradigm for photonic quantum information processing.
Yet the same physical characteristics that preserve spectral properties through fiber-optic channels also make the manipulation of frequency-encoded states in the form of quantum gates a challenge, demanding advanced optical systems that can coherently control spectral modes.
Several solutions have been explored, including nonlinear parametric mixers~\cite{Raymer2010, Kobayashi2016, Clemmen2016, Kobayashi2017, Li2019, Joshi2020b},
integrated structures based on modulated ring resonators~\cite{Zhang2019, Gevorgyan2020,Buddhiraju2021, Yuan2021,Hu2021},
and the quantum pulse gate for addressing pulsed time-frequency modes~\cite{Brecht2015, Manurkar2016, Reddy2018, Ansari2018}. 
In the case of frequency-bin encoding specifically, 
the quantum frequency processor (QFP) has been introduced and analytically shown capable of scalable universal quantum information processing~\cite{lukens16QFP,Lu2019c}.
Leveraging concatenations of alternating Fourier-transform pulse shapers and electro-optic phase modulators (EOMs)---standard components in lightwave communications---the QFP
has enabled a variety of fundamental quantum gates~\cite{Lu2018a,Lu2019a,Lu2020b,Lu2022}, supporting applications such as
quantum simulations~\cite{Lu2019b} and Bell state measurements~\cite{Lingaraju2022}.

All QFP experiments so far have relied on commercial pulse shapers and EOMs that have facilitated excellent agreement with theory (fidelity values up to $\cF=1-10^{-6}$~\cite{Lu2019b,Lingaraju2022}).
However, the use of discrete components has proven limiting in several significant ways, including spectral resolution (typically $\gtrsim$10~GHz~\cite{Roelens2008}), insertion loss ($\sim$12.5~dB for an EOM/pulse shaper/EOM QFP~\cite{Lu2018a}), and device footprint ($\sim$1~m$^2$).
The development of on-chip QFPs via integrated photonics offers potential to surpass these limitations.
In addition to the size and cost reduction of photonic integrated circuits (PICs), pulse shapers based on microring resonator (MRR) filters do not face the same resolution constraints as diffractive optics, and the elimination of fiber-coupling interfaces between components in a monolithic platform removes a major source of loss in existing QFPs.
Coupled with the rapid progress on integrated sources of high-dimensional frequency-bin-entangled states~\cite{Reimer2016, Kues2017, Imany2018, Lu2019x, Kues2019, Steiner2021, Lu2021}, integrated QFPs should enable complete frequency processing systems on chip.
Yet PIC models that incorporate the nuances of MRR pulse shapers on QFP gate performance have not been developed.

In this work, we propose a complete model for PIC-based QFPs.
Accounting for the impact of MRRs on crosstalk and dispersion, our model permits a direct mapping from geometric and material properties to quantum gate performance.
Focusing on the frequency-bin Hadamard gate in silicon as a paradigmatic example, we show that peak gate fidelity values $\cF>1-10^{-7}$ are attainable, with reasonable performance for inputs whose spectral bins fill up to $\sim$10\% of each MRR line.
We also explore the trade-off between loss and crosstalk in high-order MRR filters, in the process revealing the prospect for bin spacings as close as $\sim$1~GHz.
Our approach can be adapted to any material platform and provides an important foundation for the design of frequency-processing PICs that exceed the performance of their discrete counterparts.

\section{Model}\label{sec:model}
\subsection{Quantum Frequency Processor}\label{sec:QFP}
Optical frequency multiplexing has been used extensively by the telecom industry.
As such, components such as tunable lasers, spectrum analyzers, and electro-optic modulators are commercially available and relatively inexpensive.
The existing global optical-fiber infrastructure forms an enticing platform for scalable quantum communication and the emerging quantum internet~\cite{Wehner2018}, in turn motivating the development of cost-effective quantum-enabled devices using an optical frequency basis to process information.

Consider a collection of frequency modes centered at equispaced bins $\omega_n \triangleq \omega_0 + n\Delta\omega$ ($n\in\mathbb{Z}$), each described by a continuum of annihilation operators $\hat{a}_n(\Omega) \triangleq \hat{a}(\omega_n + \Omega)$ with $\Omega\in \left(-\frac{\Delta\omega}{2}, \frac{\Delta\omega}{2}\right)$.
In contrast to most previous QFP work~\cite{lukens16QFP}, we explicitly include dependence on the frequency offset $\Omega$, which will prove to be important when considering finite linewidth effects below.
The QFP operates on these modes with an alternating series of EOMs and line-by-line pulse shapers.
The EOMs are modeled as time-domain phase shifters that map input operators $\hat{a}_n$ to outputs $\hat{b}_m$ according to
\begin{equation}
\label{eq:EOM}
\hat{b}_m(\Omega) = \sum_{n=-\infty}^\infty c_{m-n} \hat{a}_n(\Omega)
\end{equation}
where $c_n = \frac{1}{T} \int_0^T \mathrm{d}t\, e^{i\varphi(t)} e^{in\Delta\omega t}$ and the phase function $\varphi(t)$ is periodic at the inverse mode spacing $T=\frac{2\pi}{\Delta\omega}$.
A line-by-line pulse shaper performs the transformation
\begin{equation}
\label{eq:PS}
\hat{b}_n(\Omega) = e^{i\phi_n} \hat{a}_n(\Omega).
\end{equation}

Two assumptions underpin the validity of the ideal formulae above: (i) the fundamental modulation frequency of the phase function $\varphi(t)$ is equal to the bin spacing $\Delta\omega$; (ii) the bin spacing $\Delta\omega$ is sufficiently wide, and the effective bandwidth of each bin sufficiently narrow, such that distortions in the phase mask due to the finite resolution of the pulse shaper can be neglected.
Experimentally, Assumption~(i) is typically straightforward to satisfy with a tunable radio-frequency oscillator and an initial calibration stage.
Assumption~(ii)---i.e., operation in the true line-by-line regime~\cite{Cundiff2010, Weiner2011, Torres2014}---can be built in by design.
For example, suppose that the pulse shaper resolution is such that a guardband of width $\Omega_G$ centered on the edge between adjacent bins completely encompasses the region of crosstalk and extra loss where the phase values $\phi_n$ change.
Then, by ensuring that the effective bandwidth $\Omega_B$ of each frequency bin satisfies $\Omega_B < \Delta\omega - \Omega_G$, the ideal phase expression in \cref{eq:PS} will hold for all input states~\cite{Lu2020a}.

In all QFP demonstrations so far, careful design of the physical system has ensured that \cref{eq:EOM,eq:PS} are sufficient to accurately model the results observed.
Nevertheless, the more restrictive Assumption~(ii) can be easily removed by considering the pulse shaper transformation as a generic filter $H(\omega)$ such that
\begin{equation}
\label{eq:finiteRes}
\hat{b}_n(\Omega) = H(\omega_n+\Omega) \hat{a}_n(\Omega)
\end{equation}
in lieu of pure line-by-line shaping.
For the case of a standard diffractive pulse shaper, the effects of finite spectral resolution are well known, and can be modeled by convolving an ideal stairstep phase function $H_0(\omega)$ with a resolution function $R(\omega)$: $H(\omega) = \int \mathrm{d}\omega^\prime\; H_0(\omega^\prime) R(\omega - \omega^\prime)$~\cite{Weiner2009}.

As long as \cref{eq:EOM} holds for each EOM [Assumption~(i)], only frequencies with the same offset $\Omega$ will interfere, so that the QFP can be modeled as performing a continuum of simultaneous transformations described by matrices $V(\Omega)$ such that the final output is
\begin{equation}
\label{eq:Wmatrix}
\hat{b}_m(\Omega) = \sum_{n=-\infty}^\infty V_{mn}(\Omega) \hat{a}_n(\Omega)
\end{equation}
for each offset comb defined by frequencies $\omega_n+\Omega$ ($n\in\mathbb{Z}$).
Defining the $d\times d$ matrix $W$ as the submatrix of $V$ in the logical space of interest~\cite{lukens16QFP}, we can then compute the fidelity
\begin{equation}
\label{eq:fid}
\cF_W(\Omega) = \frac{\left| \Tr \left[ W^\dagger(\Omega)U \right] \right|^2}{d^2 \cP_W(\Omega)}
\end{equation}
and success probability
\begin{equation}
\label{eq:prob}
\cP_W(\Omega) = \frac{\Tr \left[ W^\dagger(\Omega) W(\Omega) \right]}{\Tr U^\dagger U}
\end{equation}
with respect to a targeted frequency-bin unitary $U$~\cite{Uskov2009, lukens16QFP}.
The explicit $\Omega$-dependence accounts for the impact of nonuniform pulse shaper filters and will prove valuable when assessing integrated QFP designs.

Finally, before proceeding further, we note that since $|H(\omega)|\leq 1$ for a pulse shaper with finite resolution, \cref{eq:finiteRes} is in general nonunitary. This property implies that a more accurate quantum mechanical description should include mixing with vacuum modes in order to model any optical loss.
Nevertheless, in the following sections, we concentrate on discrete-variable frequency-bin encoding where it is assumed that postselection is performed on the detection of transmitted photons; in this way, any loss can be absorbed into the final success probability, permitting the use of the simpler expression in \cref{eq:finiteRes}.

\subsection{Integrated QFP}\label{sec:IQFP}
\begin{figure}[b!]
\centering\includegraphics[width=\columnwidth]{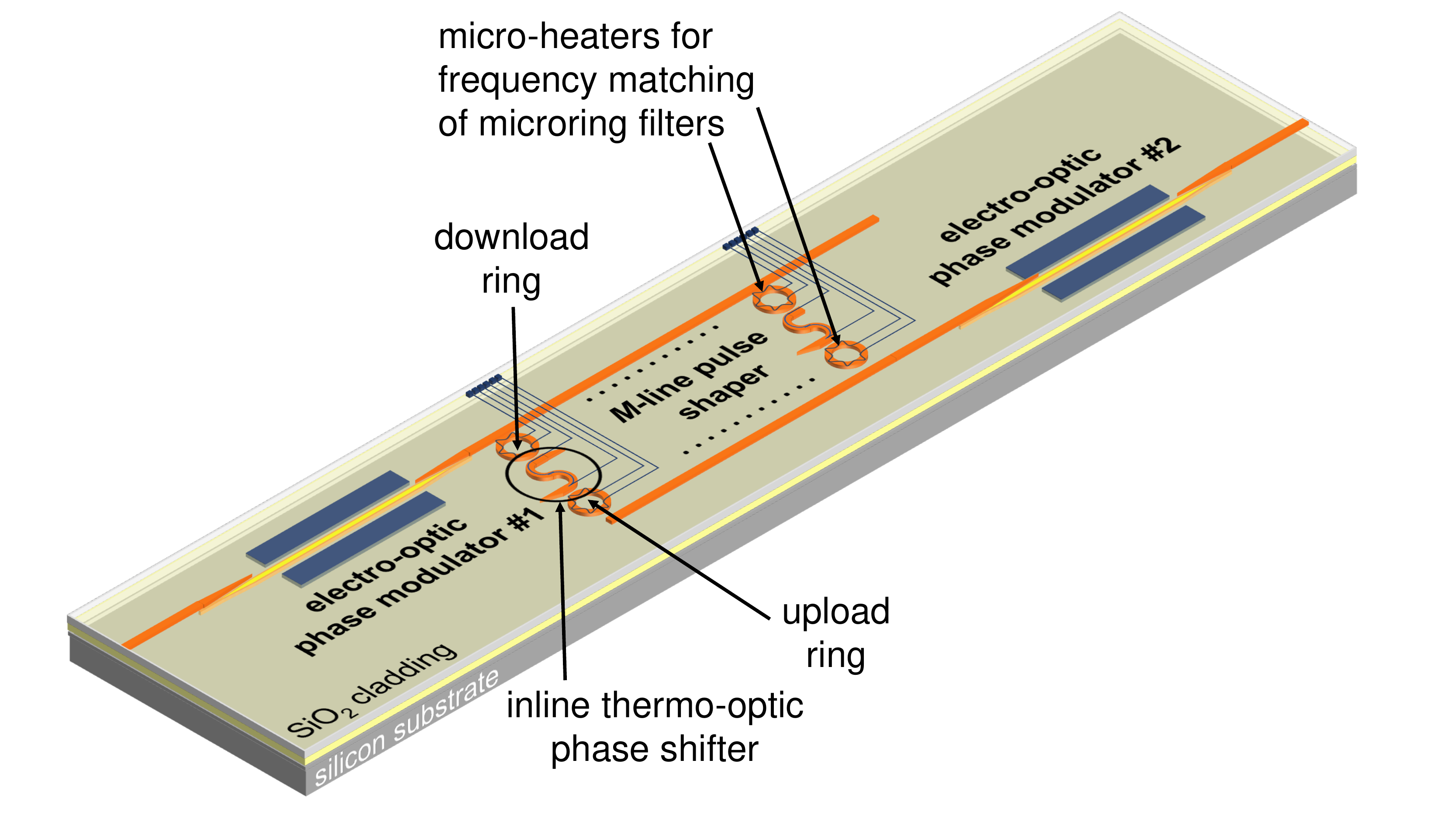}
\caption{Representative integrated QFP.
Two EOMs bookend an $M$-line MRR-based pulse shaper.
Material and geometric properties are included in the pulse shaper model, while the EOMs are treated as ideal black boxes (similar to how they are typically accessed in a process design kit).}
\label{fig:QFP}
\end{figure}
\Cref{fig:QFP} depicts our vision for an integrated QFP consisting, in this case, of two EOMs and one pulse shaper.
Following previous designs in the classical domain~\cite{Agarwal2006, Khan2010, Wang2015}, each line to be shaped is addressed and downloaded by an MRR filter, sent through a phase shifter (either thermo-optic or electro-optic), and uploaded to the common output waveguide through another MRR filter.
We define $M$ as the total number of spectral lines in the pulse shaper design, with single-ring add-drop filters ($N=1$) shown for concreteness in \cref{fig:QFP}; generalizations to $N$-ring filters are considered in \cref{sec:limits}.

The construction of high-speed and low-loss phase modulators has remained an enduring challenge in integrated photonics, complicated by the lack of an efficient electro-optic effect in silicon.
Recent developments in thin-film lithium niobate~\cite{Wang2018b, Ren2019} and CMOS-compatible silicon-organic hybrid modulators~\cite{Alloatti2014, Kieninger2018, Kieninger2020, Ummethala:21} appear particularly promising to supplant the lossy plasma-dispersion modulators standard in silicon photonics~\cite{Soref1987, Reed2010}.
Yet because modulator performance depends heavily on aspects such as material platform, doping, and device geometry---with CMOS foundries often providing black-box elements as part of their process design kits---for our purposes here we assume the availability of a linear EOM imparting the desired phase shifts without appreciable voltage-dependent loss.

\begin{figure}[tb!]
\centering
\includegraphics[width=\columnwidth]{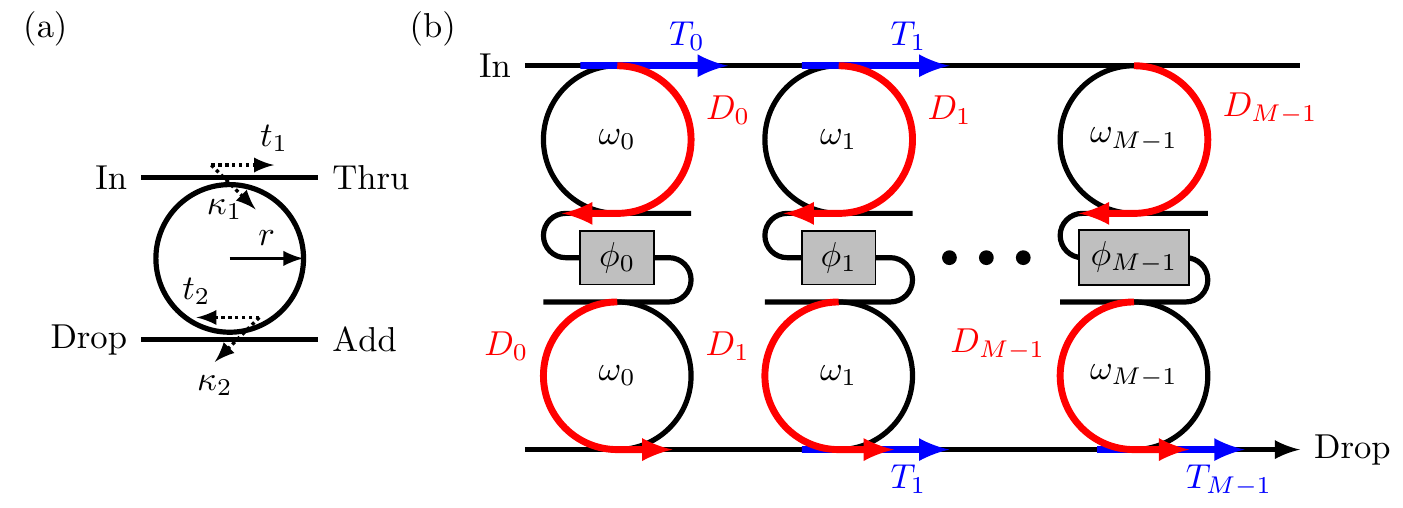}
\caption{Mathematical model of an MRR-based pulse shaper.
(a)~Parameter definitions for a single MRR.
(b)~Full pulse shaper design, where each channel $m$ is aligned to frequency $\omega_m$.
$T_m(\omega)$ and $D_m(\omega)$ describe the through and drop frequency responses of a single ring in each channel, respectively.}
\label{fig:pulseShaperDiagram}
\end{figure}
Microring resonators are important and ubiquitous components in integrated photonic devices~\cite{bogaerts11ringRes}.
A simple add-drop MRR filter is shown in \cref{fig:pulseShaperDiagram}(a) as a pair of straight parallel waveguides with an annular waveguide of radius $r$ and round-trip length $L_\text{rt}=2\pi r$ positioned between them.
As light propagates around this ring, it accumulates round-trip phase $\phi_\text{rt}(\omega)=\omega n_\text{eff}(\omega)L_\text{rt}/c$ determined by the effective index $n_\text{eff}$ of the guided optical mode in the ring and the frequency $\omega=\frac{2\pi c}{\lambda}$, with $\lambda$ the vacuum wavelength.
There is also an overall field attenuation $A=e^{-\alpha L_\text{rt}/2}$ characterized by the parameter $\alpha$ due to, e.g.,~internal scattering or absorption.
The separations between the straight and ring waveguides determine the coupling coefficients $\kappa_1$ and $t_1$ and $\kappa_2$ and $t_2$ in \cref{fig:pulseShaperDiagram}(a).
Since light may make several round trips before coupling back out, the full expressions for fields at the through ($T$) and drop ($D$) ports, relative to the input, are given by~\cite{chrostowski15photonicsDesign}
\begin{equation}
\begin{split}
    \label{eq:mrr}
    T(\omega) & \triangleq\frac{E_\text{thru}(\omega)}{E_\text{in}(\omega)}=\frac{t_1-t_2Ae^{i\phi_\text{rt}(\omega)}}{1-t_1t_2Ae^{i\phi_\text{rt}(\omega)}}\qquad\\
    D(\omega) & \triangleq\frac{E_\text{drop}(\omega)}{E_\text{in}(\omega)}=\frac{-\kappa_1\kappa_2\sqrt{A}e^{i\phi_\text{rt}(\omega)/2}}{1-t_1t_2Ae^{i\phi_\text{rt}(\omega)}}
\end{split}.
\end{equation}
Changing properties of the ring such as the radius or effective index, e.g., with an externally applied temperature or electric field gradient, shifts the resonance.

A full pulse shaper can be formed as shown in \cref{fig:pulseShaperDiagram}(b), where an array of MRRs isolate a pre-defined vector of $M$ linearly spaced frequency modes $\omega_m=\omega_0+m\Delta\omega$ ($m\in\{0,1,...,M-1\}$).
The pulse shaper then applies an arbitrary phase shift $\varphi_m$ to each filtered signal $\omega_m$ in accordance with \cref{eq:PS}.
Since light that is filtered out of the drop port of one channel must necessarily be transmitted by the through port of all other channels to reach the output, the crosstalk model for an MRR-based pulse shaper results in the discrete sum
\begin{equation}\label{eq:pulseShaper}
    H(\omega)\triangleq\frac{E_\text{drop}(\omega)}{E_\text{in}(\omega)}=\sum_{p=0}^{M-1}\left(D_p^2(\omega)e^{i\phi_p}\prod_{\substack{q=0\\q\neq p}}^{M-1}T_q(\omega)\right),
\end{equation}
in contrast to the convolution expression for a diffractive pulse shaper~\cite{Weiner2009}.
Note that this equation holds for any frequency~$\omega$, irrespective of its relationship to the designed resonances~$\omega_m$, as it accounts for all possible pathways from the input waveguide to the output.
Of course, \emph{ideally} the output at frequency $\omega\approx\omega_m$ should be dominated by the $p=m$ term, with all others negligible, but the model in no way requires this situation and therefore can incorporate all crosstalk effects automatically.
In many ways, \cref{eq:pulseShaper} represents the defining equation for MRR pulse shapers, the consequences of which are unpacked in the following sections for QFP design.

\section{Application: Hadamard Gate Performance}\label{sec:application}
\subsection{Single and Parallel Gates}\label{sec:HadamardIdeal}

For our integrated QFP simulations, we focus on the frequency-bin Hadamard gate as an archetypal example.
Not only is it a member of a universal gate set~\cite{Nielsen2000}, but the Hadamard operation---or frequency-bin beamsplitter---has also been demonstrated in multiple platforms~\cite{Clemmen2016, Lu2018a} and has facilitated observation of the frequency-bin Hong--Ou--Mandel effect~\cite{Kobayashi2016, Lu2018b}.
Moreover, an extremely attractive feature of the QFP---parallelization---has been realized experimentally for the Hadamard operation specifically~\cite{Lu2018a,Lu2018b}, thereby providing a table-top benchmark against which one can compare the integrated QFP alternative.
In the notation of \cref{sec:QFP}, the ideal mode unitary for a single Hadamard gate is 
\begin{equation}\label{eq:hadamard}
U_H=\frac{1}{\sqrt{2}}\begin{bmatrix}1&1\\1&-1\end{bmatrix},
\end{equation}
which can be designed to operate on any two frequency bins $\hat{a}_m$ and $\hat{a}_n$.

We enlist the particular Hadamard QFP solution described in Ref.~\cite{Lu2019b}, which specifies a sinusoidally varying phase modulation on each EOM with peak temporal phase deviation of 0.8283~rad and a $\pi$ stairstep phase shift between the logical zero and one modes on the pulse shaper.
In a previous experimental test with discrete components, we found that similar parallel Hadamard solutions perform with no observable crosstalk as long as they are separated by at least four guardband modes, or equivalently that each gate is assigned six unique frequency bins: two for the logical qubit space, plus two on either side as vacuum ancillas~\cite{Lu2018a}.

\begin{figure*}[tb!]
\centering
\includegraphics[width=0.8\textwidth]{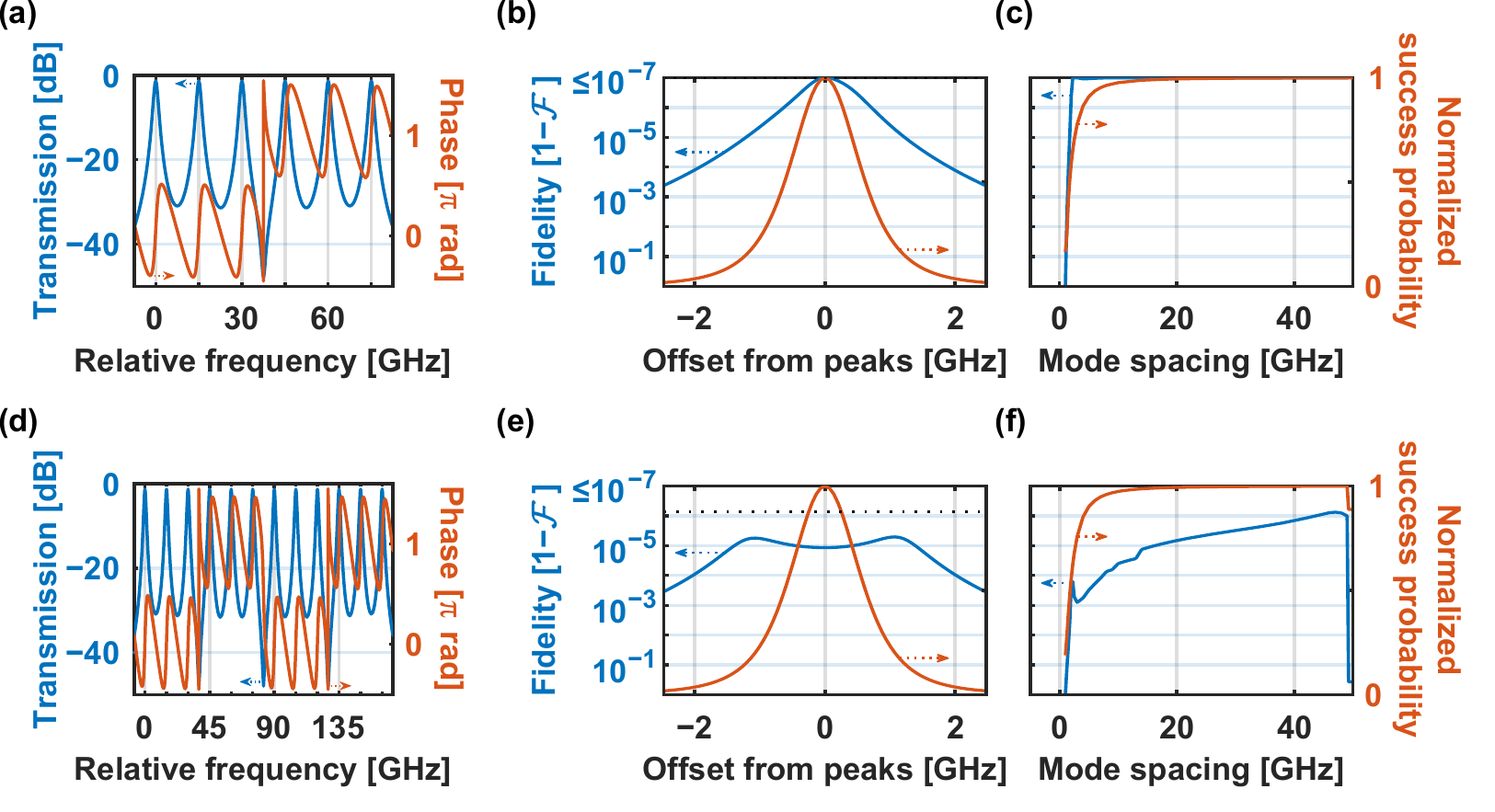}
\caption{
(a)~Transmission and phase for a pulse shaper configured for use as a Hadamard gate with $\frac{\Delta\omega}{2\pi}=15$~GHz.
(b)~Fidelity $\cF_W$ and success probability $\cP_W$ for a comb of frequency bins shifted by $\Omega$ from the transmission peaks in (a).
(c)~Fidelity $\cF_W$ and success probability $\cP_W$ for a comb of frequency bins centered at the transmission peaks ($\Omega=0$) but with variable spacing $\Delta\omega$.
(d--f)~The same as (a--c), now for a pulse shaper implementing two Hadamard gates operating in parallel.
}
\label{fig:HadamardIdealEOM}
\end{figure*}

Accordingly, in the integrated QFP designs here, we allocate $M=6$ channels for each Hadamard operation, so that, numbering channels from $m=0$, $U_H$ defined in \cref{eq:hadamard} acts on the computational modes $\hat{a}_2$ and $\hat{a}_3$.
For reference, the case of ideal linear EOMs and a perfect line-by-line pulse shaper corresponds to $\cF_W > 1-10^{-7}$ and $\cP_W=0.9745$ for this $M=6$ channel solution.
For two gates in parallel ($M=12$ channels), the Hadamard transformations are performed on pairs $(\hat{a}_2,\hat{a}_3)$ and $(\hat{a}_8,\hat{a}_9)$ from the space $\hat{a}_0$--$\hat{a}_{11}$, i.e.,
\begin{equation}\label{eq:hadamardParallel}
    U_\parallel\begin{bmatrix}
        \hat{a}_2\\
        \hat{a}_3\\
        \hat{a}_8\\
        \hat{a}_9\\
    \end{bmatrix}
    =\frac{1}{\sqrt{2}}
    \begin{bmatrix}
        1 & 1 & 0 & 0\\
        1 &-1 & 0 & 0\\
        0 & 0 & 1 & 1\\
        0 & 0 & 1 &-1\\
    \end{bmatrix}
    \begin{bmatrix}
        \hat{a}_2\\
        \hat{a}_3\\
        \hat{a}_8\\
        \hat{a}_9\\
    \end{bmatrix}.
\end{equation}

All simulations are performed in silicon at 300~K, with a fixed waveguide cross-section of 480~nm~$\times$~220~nm (width $\times$ height), for which we precompute the spectrally dependent effective index $n_\text{eff}(\omega)$ for the TE polarization using a numerical eigenmode solver.
In the interest of presenting an aggressive---yet feasible and foundry-compatible system~\cite{Carpenter2021, Bian:20}---we assume a nominal waveguide attenuation of 0.5~dB/cm ($\alpha=0.115$~cm$^{-1}$), ring radii $r\approx20\ \upmu$m, and symmetric coupling constants $\kappa_1^2=\kappa_2^2\triangleq\kappa^2=0.01$ for each ring in the pulse shaper, comfortably in the overcoupled regime ($t_1=t_2<A$) as desired for an efficient drop port response.
Further, $\frac{\omega_0}{2\pi}=193\text{ THz}$ and $\frac{\Delta\omega}{2\pi}=15\text{ GHz}$ define the nominal frequency mode space $\omega_m$ used by the pulse shaper (in the telecom band).

The success probability and fidelity are calculated via \cref{eq:fid,eq:prob} with $U=U_H$ and the mode transformation $W$ found by the model.
As shown in \cref{fig:HadamardIdealEOM}, the pulse shaper amplitude and phase varies strongly both within and across the 15~GHz-wide bins, for both a single Hadamard gate (a) and parallel Hadamard gates (d).
Such variation reiterates the need to define $\cP_W(\Omega)$ and $\cF_W(\Omega)$ with respect to frequency offset $\Omega$ from the peak.
Consequently, after adding in the mixing operations from two ideal EOMs, temporally aligned for maximum fidelity, we can compute $W(\Omega)$, and hence $\cP_W$ and $\cF_W$, not only at the predefined nominal frequencies ($\Omega=0$), but also at a series of frequency offsets relative to the mode peaks; the results are plotted in \cref{fig:HadamardIdealEOM}(b,e).

The maximum success probability for the single gate is 73.00\%, dropping slightly to 72.92\% for two parallel gates.
Note that this number includes MRR insertion loss as well as residual photon scattering into adjacent modes.
Because loss impacts coupling and linewidth of the MRR add-drop filers and is not uniform with frequency, it is not as readily distinguished from unitary scattering mechanisms as in bulk pulse shaper designs.
Consequently, comparing $\cP_W(0)=0.7300$ to the 0.9745 number expected in a lossless system offers little insight into the system's performance; since our focus lies with how $\cP_W(\Omega)$ \emph{changes} with offset, we plot success probability normalized to the peak in \cref{fig:HadamardIdealEOM}.
The fidelity remains above 0.9998 in both the single and parallel gates, even at a $\pm2\text{ GHz}$ offset from the mode peaks; the success probability is halved with an offset of $\pm0.6\text{ GHz}$.
Incidentally, were propagation loss reduced further, the peak success probability would increase accordingly: at 0.25~dB/cm, our model predicts $\cP_W(0)=0.8375$; at 0.10~dB/cm, $\cP_W(0)=0.9115$.

The plots in \cref{fig:HadamardIdealEOM}(c,f) show the fidelity and success probability at offset $\Omega=0$, but now as a function of mode spacing $\frac{\Delta\omega}{2\pi}$.
Increasing the mode spacing initially reduces the crosstalk noise between neighboring modes, leading to steady increases in $\cF_W$ and $\cP_W$.
Yet in the parallel case, this trend is not monotonic, with an initial fidelity maximum at 2.25~GHz beyond which the fidelity decreases briefly, before rising more gradually over the explored domain to a maximum around $\sim$47~GHz, and finally crashing down to $\cF_W<0.5$ at mode spacings beyond $\sim$49~GHz.

Ostensibly, such behavior seems quite surprising: why should fidelity \emph{decrease} at greater mode separations?
Yet this phenomenon can be traced to the periodicity of the MRR transfer functions.
For the current radii $r\approx20\ \upmu$m (each slightly adjusted to align with a specific resonance $\omega_m$), the free spectral range (FSR) is approximately 564.5~GHz.
Thus, if the total pulse shaper bandwidth $M\frac{\Delta\omega}{2\pi}$ is well below $564.5$~GHz, the effects of additional resonances adjacent to the band of interest are negligible.
Indeed, for the case of the parallel Hadamard gates with a total $M=12$ channels, when the channels fill a single FSR exactly---i.e., $\frac{\Delta\omega}{2\pi}=564.5\text{ GHz}/12=47.0$~GHz---\cref{fig:HadamardIdealEOM}(f) shows a peak in fidelity corresponding to the minimum possible channel crosstalk.
Immediately beyond this point, the fidelity decreases precipitously as $\omega_M$ begins to undesirably couple to the next resonance of the $\omega_0$ ring, and vice versa.
By contrast, since the single Hadamard gate uses only $M=6$ pulse shaper channels, it remains comfortably away from this FSR limit and shows no such drop on the scale of \cref{fig:HadamardIdealEOM}(c).


\subsection{Broadband Performance}\label{sec:Broadband Preformance}
The effects of frequency offset on fidelity and success probability are evident in \cref{fig:HadamardIdealEOM}(b,e).
However, this frequency-offset analysis alone is not sufficient to fully predict the performance of the QFP for arbitrary, pulsed inputs, which cannot be viewed as a frequency comb composed of infinitesimally narrow lines, but rather consist of spectral modes comprising a continuum of frequencies occupying some fraction of the MRR resonance.
Within a group of equispaced, infinitesimally narrow inputs (the case in \cref{fig:HadamardIdealEOM}), any overall output phase shift has no impact on the computed fidelity.
In contrast, for broadband inputs, residual phase shifts that vary with offset will affect the overall spectro-temporal shape of the output, and must be accounted for.
For example, while the spectral phases applied by the pulse shapers in \cref{fig:HadamardIdealEOM}(a,d) appear roughly constant at any \emph{given} offset $\Omega$, they vary strongly \emph{between} offsets; therefore it is critical to quantify what impact this filtering may have on the shape of finite-linewidth inputs.

\begin{figure}[tb!]
\centering
\includegraphics[width=\columnwidth]{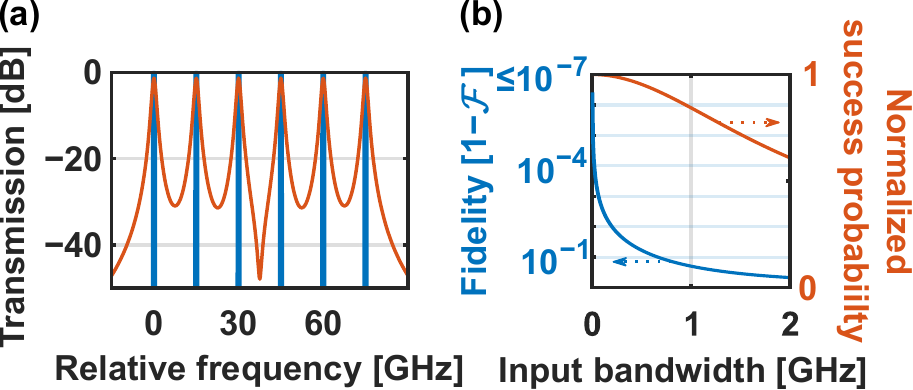}
\caption{(a)~Example Nyquist input spectrum compared to typical pulse shaper filter.
The rectangular input modes are 0.97~GHz wide, equal to the half-width at half-maximum bandwidth of each pulse shaper channel.
(b)~Fidelity $\cF_{y}$ and success probability $\cP_{y}$ vs.~input bandwidth for input $\ket{+}$ into a QFP Hadamard gate.}
\label{fig:Broadband}
\end{figure}

For simulation purposes, we select tunable-bandwidth inputs that satisfy the Nyquist criterion~\cite{Nakazawa2012,Soto2013,Leuthold2015}, specifically 
perfect sinc pulses of the form
\begin{equation}
s(t) = \frac{\sin\frac{\pi t}{T_{s}}}{\frac{\pi t}{T_{s}}},
\label{eq:pulseSinc}
\end{equation}
corresponding to an ideal rectangular spectrum $S(\omega)$ tunable in width by varying the symbol frequency $\frac{1}{T_s}$:
\begin{equation}
S(\omega) = \begin{cases}
\sqrt{\frac{T_s}{2\pi}} & |\omega|<\frac{\pi}{T_s} \\
0 & |\omega|>\frac{\pi}{T_s}
\end{cases},
\label{eq:pulseSpectrum}
\end{equation}
normalized such that $\int\mathrm{d}\omega\, |S(\omega)|^2=1$.
Although highly idealized, selection of this spectral shape is particularly convenient for bandwidth analyses, allowing us to probe the frequency response of the device under test while minimizing the potential for artifacts from features in the probe signal like spectral tails. 
\Cref{fig:Broadband}(a) depicts an example of the frequency comb formed by defining rectangular frequency bins according to \cref{eq:pulseSpectrum}.
Considering computational modes $(\omega_2,\omega_3)$ for the $M=6$ Hadamard gate, we can encode a generic input qubit $\ket{\psi}=c_0\ket{0}+c_1\ket{1}$ as a single-photon wavepacket with complex spectrum $x(\omega)=c_0 S(\omega-\omega_2) + c_1 S(\omega-\omega_3)$.

The equations used to calculate the fidelity and success probability in the previous sections assume that the shape of the input modes is unchanged by the operation.
The output wavepacket would be $g(\omega)=\tilde{c}_0 S(\omega-\omega_2) + \tilde{c}_1 S(\omega-\omega_3)$, where the new coefficients are as determined by the Hadamard gate: $[\tilde{c}_0\, \tilde{c}_1]^T = U_H [c_0 \,c_1]^T$.
This assumption does not hold for the broadband input model, as only a portion of the signal is located at each channel peak [cf.~\cref{fig:Broadband}(a)].
The mode shapes $S(\omega)$ are themselves modulated by the QFP, and therefore in this section we define fidelity and success probability for the actual output wavepacket $y(\omega)$ as
\begin{equation}
\label{eq:fid_Broadband}
    \cF_{y}=\frac{\left|\int\mathrm{d}\omega\, g^*(\omega)y(\omega)\right|^{2}}{\int\mathrm{d}\omega\, \left|g(\omega)\right|^2 \int\mathrm{d}\omega\, \left|y(\omega)\right|^2 }
\end{equation}
and
\begin{equation}\label{eq:prob_Broadband}
    \cP_{y}= \frac{\int\mathrm{d}\omega\, \left|y(\omega)\right|^2}{\int\mathrm{d}\omega\, \left|g(\omega)\right|^2},
\end{equation}
where $g(\omega)$ and $y(\omega)$ are the outputs of a finite-bandwidth input that has undergone an ideal $U_H$ and actual $W$ transformation, respectively.

In \cref{fig:Broadband}(b) are the Hadamard gate $\cF_{y}$ and $\cP_{y}$ results for the input state $\ket{+} = \frac{1}{\sqrt{2}}(\ket{0} + \ket{1})$, with full bandwidths ranging from 0--2~GHz, for a bin spacing $\frac{\Delta \omega}{2\pi} = 15~\textrm{GHz}$.
Note that, because it is possible for the change in spectral mode shapes to vary with input state, $\cF_y$ and $\cP_y$ are defined for a specific input-output combination; in practice, we noticed no difference in bandwidth scaling for all states examined, so we plot only the results for the $\ket{+}$ input for simplicity here.
As expected, these broadband inputs prove significantly more sensitive to phase variations in the pulse shaper spectrum than the offset tests of \cref{sec:HadamardIdeal}.
Whereas the fidelity values in \cref{fig:HadamardIdealEOM}(b)
remain above 0.999 over the entire peak, \cref{fig:Broadband}(b) shows that an input less than $\sim$60~MHz wide is required to retain the same fidelity in the pulsed case.
Nonetheless, $\cF_y>0.99$ holds up to an input bandwidth $\sim$10\% of the total MRR linewidth, an appreciable fraction for which high fidelity values should be possible experimentally.

\section{Pushing Device Limits: High-Order Filtering}\label{sec:limits}
To maximize spectral efficiency one would like to operate at as small a mode spacing $\Delta\omega$ as possible.
Yet as revealed in \cref{fig:HadamardIdealEOM}(c,e), success probability begins to drop for $\frac{\Delta\omega}{2 \pi}\lesssim10$~GHz for the standard single-ring filter design.
It is the goal of this section to explore the prospect for much tighter spacings with higher-order MRR filters.
Coupled MRRs have have been shown to give transmission passbands a sharper rolloff and wider bandwidth~\cite{little1997microring}, and have been used to increase spectral efficiency, for example in the generation of spectral-phase-encoded signals for optical code-division multiple-access~\cite{Agarwal2006}.
Thus, by the addition of coupled MRRs to the pulse shaper design in \cref{fig:pulseShaperDiagram}, one should be able to bring the frequency bins closer together without interchannel crosstalk. 
Extending our model to include coupled MRRs can be realized with the transfer matrix method~\cite{poon2004matrix}; this modifies the through $T(\omega)$ and drop $D(\omega)$ functions in \cref{eq:mrr}, but otherwise the calculations proceed unchanged.
As above, we take the coupling constant between each bus waveguide and adjacent MRR to be $\kappa^{2} = 0.01$, selecting the inter-MRR couplings according to the pattern of values in Table~1 of Ref.~\cite{little1997microring} to achieve flat passbands with strong rolloff.

\begin{figure}[tb!]
\centering
\includegraphics[width=\columnwidth]{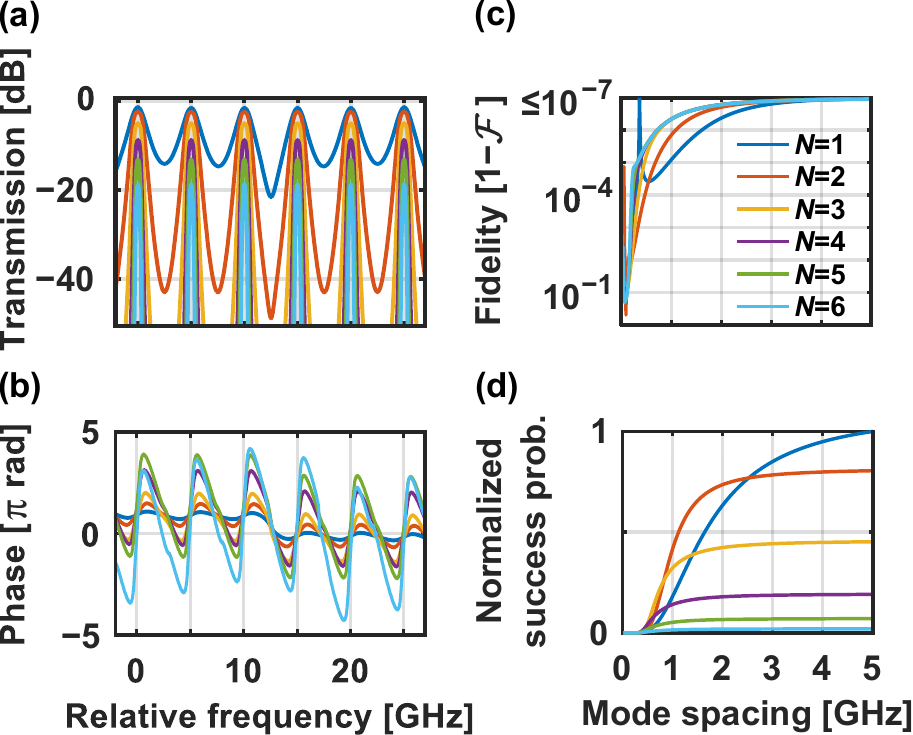}
\caption{$N$-order filter performance (up to $N = 6$ coupled MRRs):
(a)~Pulse shaper drop port transmission spectrum for a QFP with 5~GHz mode spacing.
(b)~Pulse shaper drop port phase spectrum for a QFP with 5~GHz mode spacing.
(c)~$\cF_W(0)$ vs.~mode spacing and (d)~$\cP_W(0)$ (normalized) vs.~mode spacing for the corresponding peaks in (a).}
\label{fig:N_MRR Results}
\end{figure}

\Cref{fig:N_MRR Results}(a,b) provides the pulse shaper drop port transmission and phase spectra for an $M=6$ Hadamard gate at spacing $\frac{\Delta\omega}{2\pi} = 5\text{ GHz}$ and up to ($N=6$)-order filters.
The isolation between adjacent bins increases rapidly with $N$, as designed.
Yet because each channel path includes a total of $2N$ MRRs ($N$ for drop, $N$ for add), the 0.5~dB/cm waveguide loss leads to significant drops in peak transmission.
Both of these effects appear in the zero-offset modal fidelity $\cF_{W}(0)$ and success probability $\cP_{W}(0)$, plotted in \cref{fig:N_MRR Results}(c,d).
Importantly, the higher-order filters do succeed in sustaining high fidelity values well below 1~GHz mode spacing.
Unfortunately, though, the success probabilities also experience strong reductions in their overall value, indicating that the advantages gained from lower crosstalk are counteracted in part by reduced throughput.
By plotting all probabilities normalized to the $N=1$ configuration at 5~GHz in \cref{fig:N_MRR Results}(d), we are able to see the crossover points at which high-order filters gain in absolute throughput.
For example, $\cP_W$ for $N=2$ is seen to exceed that for $N=1$ for $\frac{\Delta\omega}{2\pi}\lesssim 2.5$~GHz; similar intersections can be found for any other two designs.
At this waveguide loss, it certainly seems to us that high-order filters beyond $N=3$ or so provide diminishing returns, exemplifying the critical place of loss in designing integrated QFPs: improving the efficiency of waveguide propagation can alter significantly which pulse shaper design will optimize performance in a particular application.

As an interesting aside, we note an unexpected resurgence in fidelity that occurs around 0.4~GHz mode spacing for the QFP design with $N=1$ filters in \cref{fig:N_MRR Results}(c).
Upon deeper investigation, we found that despite the increased presence of passband overlap, our design does indeed produce gates that more closely resemble the true Hadamard transformation at 0.4~GHz, due to an auspicious combination of crosstalk that actually improves the fidelity for this particular case.
However, the low success probability for the $N=1$ design at $\sim$0.4~GHz suggests that the temporary uptick in fidelity is more of a curiosity than an effect of practical value.

\section{Discussion and Conclusion}\label{sec:discussion}
We have shown here that integrated QFPs are feasible under realistic manufacturing constraints.
Nonetheless, our developments are not without a few key simplifications.
For example, we have assumed the availability of ideal, linear EOMs.
As each foundry typically has its own proprietary designs for highly linear analog modulators, specializing to any particular EOM design would have distracted from the main questions of the paper.
However, we do discuss one method for modeling nonlinearities in Appendix~\labelcref{sec:eom} for the case of a plasma dispersion silicon modulator~\cite{Soref1987, Reed2010, chrostowski15photonicsDesign}, an approach which can be adapted as needed to whatever modulator technology is available.

We have also assumed that thermal crosstalk between nearby phase shifters and heaters is negligible so that each MRR can be tuned independently into its desired resonance.
Efforts to minimize crosstalk would again depend on the particular geometry and material, and recent tuning algorithms appear quite promising to obtain desired filter characteristics even with strong heater-heater coupling~\cite{milanizadeh19thermal}.
In addition, undercut heaters promise high thermal isolation and significantly reduced thermal tuning power requirements~\cite{Giewont2019}.

While we have focused on silicon as an integrated photonic platform, our procedures are general, and may be readily extended to other promising photonic materials including III-V semiconductors, silicon carbide, and thin-film lithium niobate~\cite{Moody2022}.
Indeed, the exceptional efficiency and speed of electro-optic modulators in some of these materials---particularly lithium niobate~\cite{Wang2018b}---suggests that platforms other than silicon may ultimately prove better suited to the demands of the integrated QFP anyway.
Irrespective of technical limitations from material properties or questions of thermal engineering, the MRR pulse shaper relation [\cref{eq:pulseShaper}] provides fundamental limits to QFP design that should apply to any future devices based on such pervasive filter technology.

On the quantum side, we selected the Hadamard gate as the simplest nontrivial QFP operation which is also part of a minimal gate set for universal quantum computation~\cite{Nielsen2000}.
Future work could explore implementations of more complicated or indeed multiphoton gates, such as the frequency-bin controlled-NOT~\cite{Lu2019a}.
In general, we anticipate behavior with similar qualitative features to that found here: high fidelity for narrow frequency bins centered on each filter peak, which reduces as either the bins are widened or detuned.
Of course, the specific tolerances to retain a desired fidelity will no doubt vary with gate, depending on aspects such as the number of frequency bins shaped, the spectral phase patterns implemented, and the number of EOMs and pulse shapers; in any case, design simulations should be implemented specialized to the selected operation.

In fact, pushing even further and extending our analysis to the multiphoton extreme of continuous-variable (CV) encoding~\cite{Braunstein2005a, Weedbrook2012, Pfister2020} should provide valuable opportunities for QFPs specifically designed for emerging applications in CV photonics, such as Gaussian boson sampling~\cite{Hamilton2017} and Gottesman--Knill--Preskill qubits~\cite{Gottesman2001}.
The development of integrated CV sources~\cite{Paesani2019, Arrazola2021} and theory showing the QFP's potential in non-Gaussian state engineering~\cite{Pizzimenti2021} support a promising outlook for monolithic platforms combining squeezed-state generation and QFP-based control.
Nevertheless, significant enhancements to the model presented here will be required.
In particular, expressions such as \cref{eq:finiteRes}---which does not preserve commutation relations---or \cref{eq:fid_Broadband,eq:prob_Broadband}---which assume postselection on a single photon in defining wavepacket $y(\omega)$---must be generalized to treat loss in a fully quantum mechanical fashion, possible in light of recent CV formalisms~\cite{Gagatsos2019,Quesada2019}.

\section*{Acknowledgments}
We thank A.~M. Weiner and K.~V. Myilswamy for valuable discussions.
This research was performed in part at Oak Ridge National Laboratory, managed by UT-Battelle, LLC, for the U.S.~Department of Energy under contract no.~DE-AC05-00OR22725. This work was funded by the U.S. Department of Energy, Office of Science, Office of Workforce Development for Teachers and Scientists Science Undergraduate Laboratory Internship Program; the U.S. Department of Energy, Office of Science, Advanced Scientific Computing Research, Early Career Research Program (Field Work Proposal ERKJ353); the National Science Foundation (1747426-DMR, 1839191-ECCS); and Air Force Research Laboratory (FA8750-20-P-1705). This material is based upon work partially supported by the U.S.~Department of Energy Office of Science National Quantum Information Science Research Centers.


\appendices

\section{EOM Nonidealities}\label{sec:eom}
\begin{figure}[tb!]
\centering
\includegraphics[width=\columnwidth]{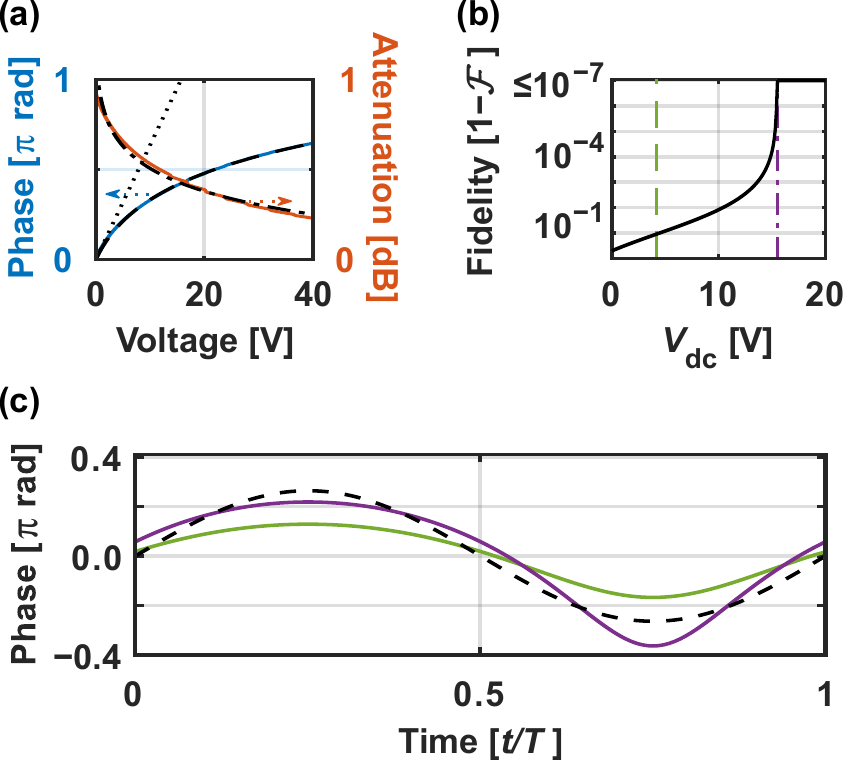}
\caption{(a)~Example nonlinear EOM response.
A linear approximation tangent to the curve at 0~V is shown for comparison.
(b)~Fidelity obtained by optimizing $V_1$ in \cref{eq:eomVoltage} for a range of $V_\text{dc}$ values.
(c)~Temporal phase applied over one cycle of \cref{eq:eomVoltage} for the points highlighted in (b), compared to the targeted sinewave (dotted).}
\label{fig:EOM}
\end{figure}

Whereas all analyses in the main text assume an EOM with an ideal linear response, recognizing and modeling the true nonlinear behavior of a silicon EOM can permit a more realistic prediction of Hadamard gate performance, as well as reveal methods to compensate it.
\Cref{fig:EOM}(a) shows one model using the same geometry, index, and doping parameters described in Ch.~6 of Chrostowski and Hochberg's book~\cite{chrostowski15photonicsDesign} for the phase shift and attenuation resulting from a 1~mm-long EOM as functions of the applied (reverse) voltage.

We fit the phase shift curve to a logarithmic function
\begin{equation}\label{eq:eomPhase}
    \Delta\phi=a\ln\left(1+\frac{V}{V_0}\right),
\end{equation}
where in the QFP solution the drive voltage $V$ is sinusoidal:
\begin{equation}\label{eq:eomVoltage}
    V=V_\text{dc}+V_1\sin\Delta\omega t.
\end{equation}
In addition, we plot in \cref{fig:EOM}(a) the small-signal linear approximation of the phase shift, $\Delta\phi\approx \frac{aV}{V_0}$, for the sake of comparison with the function itself.
For a given bias voltage $V_\text{dc}$, we find the modulation amplitude $V_1$ which maximizes the QFP gate fidelity for a single Hadamard operation, subject to the restriction $V_1\leq V_\text{dc}$ (so that the overall reverse voltage $V\geq0$ at all times).
For these tests, we take the pulse shaper as ideal, focusing only on the EOM-induced effects.

The fidelity values found in this way are plotted in \cref{fig:EOM}(b).
The maximum fidelity increases steadily until leveling off around $V_\text{dc}\approx 16$~V, which can be explained from the $V>0$ restriction.
Below 16~V bias, the optimal oscillating voltage is precisely at the limit $V_1=V_\text{dc}$; thus, fidelity improves with increasing $V_\text{dc}$ as $V_1$ can be made larger.
On the other hand, beyond $V_\text{dc}\approx 16$~V, $V_1<V_\text{dc}$ for optimal fidelity and is thus no longer limited by this physical constraint, so attainable fidelity ceases to improve significantly with larger $V_\text{dc}$.
The purple dash-dotted line shows a high-performance operating point at $V_1=V_\text{dc}=15.5$~V with $\cF_W=1-10^{-7}$.

As a comparison, we can also use the linear approximation of the applied phase shift to determine the amplitude $V_1$ expected to give the desired peak phase shift $\Delta\phi=0.8283$.
This emulates the situation in which the QFP designer assumes ideal modulator characteristics in selecting the drive voltage, when in fact nonlinearity of the actual device cannot be neglected.
This linear approximation predicts an ideal amplitude $V_1=V_\text{dc}=4.14$~V, which in reality leads to $\cF_W<0.9$ as can be seen at the green dashed line \cref{fig:EOM}(b).
To better understand how this situation obtains, \cref{fig:EOM}(c) plots the actual temporal phases over a single period for both input voltages highlighted in (b).
The voltage recommended by the linear approximation design yields phase shifts significantly smaller than the targeted sinewave, as it overestimates the phase shift imparted, while the fully nonlinear design recognizes and attempts to compensate for the saturated response, yielding a waveform much closer to the ideal.

Finally, it is important to note that the model described here is by no means limiting; the true model for a physical device will depend on where and how it is manufactured, as well as on various design parameters.
In particular the logarithmic model for the phase modulator as described in \cref{fig:EOM} is also only an approximation which could potentially be adjusted to achieve better performance.
For example, drawing on work from the linearization of Mach--Zehnder modulators~\cite{Khilo:11}, one might look to cancel the nonlinear refractive index response with the optical loss response through careful choice of the device length, potentially requiring vertically doped junctions to maximize overlap of the optical mode with regions of index change.

\end{document}